\documentclass[conference]{IEEEtran}

\usepackage{amsmath}
\usepackage{amsfonts}
\usepackage{amssymb}
\usepackage{makeidx}
\usepackage{graphicx}
\usepackage{epstopdf}
\usepackage{cite}
\usepackage{paralist}

\usepackage{fancyhdr}
\usepackage{graphicx, amsfonts,amsmath, amssymb,array,url}
\usepackage{color}
\usepackage[usenames,dvipsnames,svgnames,table]{xcolor}
\usepackage{epstopdf}
\usepackage{graphicx}
\usepackage{amssymb}
\usepackage{amsmath}
\usepackage{cite}
\usepackage{subfigure}
\usepackage{mathrsfs}
\usepackage{tabulary}
\usepackage{multirow}
\usepackage[displaymath,mathlines]{lineno}

\ifCLASSINFOpdf

 \newtheorem{proposition}{Proposition}

\else

\fi

\newcommand{\pu}{P_{\text{u}}}
\newcommand{\pr}{P_{\text{r}}}

\newcommand{\G}{\mathbf{G}}
\newcommand{\D}{\mathbf{D}}

\newcommand{\g}{\mathbf{g}}

\newcommand{\yr}{\mathbf{y}_{\tt{R}}}

\newcommand{\B}[1]{\mathbf{#1}}

\newcommand{\CN}{\mathcal{CN}}
\newcommand{\V}{\mathcal{V}}
\newcommand{\Vt}{\mathcal{V}_{k,t'}}
\newcommand{\nr}{\mathbf{n}_{\tt{R}}}

\begin{document}

\title{How to Scale Up the Spectral Efficiency of Multi-way Massive MIMO Relaying?}

\author{\IEEEauthorblockN{Chung Duc Ho\IEEEauthorrefmark{1},
Hien Quoc Ngo\IEEEauthorrefmark{01}\IEEEauthorrefmark{2},
Michail Matthaiou\IEEEauthorrefmark{1}, and Trung Q. Duong\IEEEauthorrefmark{1}}
\IEEEauthorblockA{\IEEEauthorrefmark{1}School of Electronics, Electrical Engineering and Computer Science, Queen's University Belfast, BT7 1NN, Belfast, U.K.}
\IEEEauthorblockA{\IEEEauthorrefmark{2}Department of Electrical Engineering (ISY),
Link\"{o}ping University, 581 83 Link\"{o}ping, Sweden} 
\IEEEauthorblockA{\IEEEauthorrefmark{0}Email:\{choduc01, m.matthaiou, trung.q.duong\}@qub.ac.uk, hien.ngo@liu.se}}

\maketitle

\begin{abstract}
This paper considers a decode-and-forward (DF) multi-way massive multiple-input multiple-output (MIMO) relay system where many users exchange their data with the aid of a relay station equipped with a massive antenna array. We propose a new transmission protocol which leverages successive cancelation decoding and zero-forcing (ZF) at the users. By using  properties of massive MIMO, a tight analytical approximation of the spectral efficiency is derived. We show that our proposed scheme uses only half of the time-slots required in the conventional scheme (in which the number of time-slots is equal to the number of users \cite{ho2017performance}), to exchange data across different users. As a result, the sum spectral efficiency of our proposed scheme is nearly double the one of the conventional scheme, thereby boosting the performance of multi-way massive MIMO to unprecedented levels.

\end{abstract}


\IEEEpeerreviewmaketitle

\vspace{0.2cm}
\section{Introduction}

In the past few years, massive MIMO technology has attracted significant research attention for its ability to  improve the spectral and energy efficiency \cite{larsson2014massive,ngo2013energy}. In massive MIMO systems, many users can be served by a base station equipped with very large antenna arrays. With very large antenna arrays at the base station, the channels between different users become pairwise orthogonal, and hence, the noise and inter-user interference reduce noticeably without improving the complexity of the system \cite{ngo2013energy}. Furthermore, by using time division duplex (TDD) mode, the channel estimation overhead depends only on the number of active users regardless of number of base station antennas \cite{marzetta2016fundamentals}. This makes massive MIMO scalable and become one of the key candidates for future wireless communication systems.

On a parallel avenue, multi-way relaying networks have also been investigated to enhance the robustness against the channel variations in distinguished areas, where the direct channels among users are unavailable due to large obstacle and/or heavy path loss in the propagation environment \cite{gunduz2013multiway}. With the help of the relay station, users that are geographically separated can communicate or exchange their data-bearing symbols much easier. Moreover, a significant number of papers demonstrate that multi-way relaying networks provide much higher spectral efficiency  and communication reliability compared to one-way or two-way relaying systems \cite{tian2014degrees,amah2009non}.

The combination of multi-way relaying and massive MIMO  is very promising  since it reaps all benefits of both technologies. Recently, some papers have evaluated the performance of multi-way relaying networks with massive arrays at the relay   \cite{badugewireless, amarasuriya2014multi}. In these works, the authors showed that multi-way massive MIMO relay systems can offer huge  spectral and energy efficiency. In addition, by using simple linear  processing (e.g. ZF and maximum ratio processing)  and employing a large number of antennas at the relay station, the transmit power of each user can be scaled down proportionally to the number of relay antennas, while maintaining a given quality of service.  However, all of aforementioned studies considered a conventional transmission protocol which requires $K$ time-slots to exchange data among $K$ users.

Different with previous works, in this paper we propose a novel transmission protocol for multi-way massive MIMO relay networks which requires only $\left \lceil \frac{K-1}{2} \right \rceil+1$ time-slots for the information exchange among the $K$ users. We consider the DF operation at the relay, and assume that the relay and the users have perfect knowledge of the channel state information (CSI). We derive an approximate closed-form expression for the spectral efficiency. The approximation is shown to be very tight, especially when the number of relay antennas is large.

\textit{Notations:}  Matrices and vectors are expressed as upper and lower case boldface letter, respectively. The superscripts $(\cdot)^H$ and $\text{Tr}(\cdot)$ stand for Hermitian transpose and the trace, respectively. We denote by $\B{a}_{k}$ the $k$-th column of matrix $\B{A}$. The symbol $\|\cdot\|$ indicates the norm of a vector. The notation $\mathbb{E}\{\cdot\}$ is the expectation operator. The notation $\left[\B{A}\right]_{mn}$ or ${a}_{mn}$ denotes the $(m,n)$-th element of matrix $\B{A}$, and $\B{I}_{K}$ is the $K\times K$ identity matrix.

\section{System model}

We consider a DF  multi-way relay networks with a very large antenna array at the relay station. The system includes one relay station equipped with $M$ antennas and $K$ single-antenna users. The bearing-messages from $K$ users are exchanged with the help of the relay station. Each user wants to detect the signals transmitted from $K-1$ other users.
We assume that the users and the relay station operate in half-duplex mode and know perfectly CSI.  Furthermore, we assume that the direct links (user-to-user links) are unavailable due to large path loss and/or severe shadowing. 

The channel matrix between the $K$ users and $M$ antennas at the relay is denoted by $\G\in\mathbb{C}^{M\times K}$ and is modeled as
\begin{align}\label{eq:G}
\G=\B{H}\D^{1/2},
\end{align}
where $\B{H}\in\mathbb{C}^{M\times K}$ models small-scale fading with independent $\CN(0,1)$ components, and $\D\in~\mathbb{C}^{K\times K}$ is the diagonal matrix of large-scale fading (path loss and log-normal attenuation). Let $g_{mk}$ and $h_{mk}$ be the $(m,k)$-th element of $\G$ and $\B{H}$, respectively. Then
\begin{align}\label{eq:g_mk}
g_{mk}=\sqrt{\beta_{k}}h_{mk},
\end{align}
where $\beta_k$ is the $k$-th diagonal element of $\D$. In general, the transmission protocol is divided into two phases: multiple-access phase and broadcast phase. In the multiple-access phase,  all $K$ users transmit signals to the relay station.  In the broadcast phase, the relay station broadcast signals (which are decoded in the multi-access phase) to the users. In the next sections, we will first present the conventional transmission protocol, followed by the proposed transmission scheme.

\vspace{0.2cm}
\section{Conventional Transmission Protocol}\label{sec:convt}
In this section, we first summarize a conventional transmission protocol tailored to multi-way massive DF relaying networks. The uplink and downlink spectral efficiencies are then provided in closed-form.

\subsection{Multiple-Access Phase}\label{sec:conv_MA}

This phase requires only one time-slot. All the $K$ users transmit their data to the relay in the same time-frequency resource. The $M\times 1$ received vector at the relay is 
\begin{align}\label{eq:yr}
\yr=\sqrt{\pu}\G\B{x} + \nr,
\end{align}
where $\B{x}\triangleq \left[x_1,x_2\dots,x_K\right]^T$ is the signal vector transmitted from the $K$ users, with $\mathbb{E}\left\{\B{x}\B{x}^H\right\}=\B{I}_K$, $\nr$ is the noise vector with  i.i.d. $\CN(0,1)$  elements, and $\pu$ is the normalized transmit power of each user.

After receiving the transmitted signals from the $K$ users, the relay employs maximum ratio combining scheme by multiplying $\yr$ with $\G^H$ as follows:
\begin{align}\label{eq:r}
\B{r}=\G^H\yr
&=\sqrt{\pu}\G^H\G\B{x}+\G^H\nr.
\end{align}
Then, the $k$-th element of $\B{r}$, denoted by $r_{k}$,  is used to decode the signal transmitted from user $k$. From \eqref{eq:r}, $r_{k}$ is given by 
\begin{align}\label{eq:r_k}
r_{k}
=\sqrt{\pu}\|\g_{k}\|^2x_{k}+\sqrt{\pu}\sum\limits_{\underset{i\neq k}{i=1}}^{K}\g^H_{k}\g_{i}x_{i} +\g^H_{k}\nr,
\end{align}
where $\g_{k}$ is the $k$-th column of $\G$. Therefore, the uplink spectral efficiency of the system in \eqref{eq:r_k} (measured in bit/s/Hz) is given by 
\begin{align}\label{eq:R_ul}
{\tt{R}}^{\text{ul}}_{k}
=\mathbb{E}\left\{\log_{2}\left(1+\frac{\pu\|\g_{k}\|^4}{\pu\sum\limits_{\underset{i\neq k}{i=1}}^{K}\left|\g^{H}_{k}\g_{i}\right|^2+\|\g_{k}\|^2}\right)\right\}.
\end{align}

By using Jensen's inequality, a closed-form expression lower bound of the spectral efficiency  \eqref{eq:R_ul} is given by \cite[Eq.~(16)]{ngo2013energy}
\begin{align}\label{eq:R_tilde_ul}
{{\tt{R}}}^{\text{ul}}_{k}
\geq\tilde{{\tt{R}}}^{\text{ul}}_{k}=\log_{2}\left(1+\frac{\pu(M-1)\beta_{k}}{\pu\sum_{\underset{i\neq k}{i=1}}^{K}\beta_{i}+1}\right).
\end{align}

\vspace{-0.2cm}
\subsection{Broadcast Phase}\label{sec:conven_broads}
In this phase, the relay station transmits all signals decoded in the multiple-access phase to all users in $K - 1$ time slots. In the $t$ time-slot, the relay aims to transmit $x_{j(k,t)}$ to user $k$, $k=1, \ldots, K$, where 
\begin{align}\label{eq:j_kt}
j(k,t)\triangleq
\left
\{\begin{matrix}
(k+t)\ \text{modulo}\ K,  &\ \text{if} \ \ (k+t)\neq K \\ 
K,& \ \ \ \text{otherwise}.
\end{matrix}
\right.
\end{align}
More precisely, in the $t$-th time-slot, the relay station transmits 
\begin{align}
\B{s}^{(t)}
&=\sqrt{\frac{\pr}{M\sum_{i=1}^{K}\beta_{i}}}\sum_{i=1}^{K}\g_{i}x_{j(i,t)},\label{eq:s_t}
\end{align}
where $\pr$ is the normalized transmit power at the relay. Then,  the received signal at the $k$-th user is
\begin{align}
y^{(t)}_{k}
&=\g^H_{k}\B{s}^{(t)}+n^{(t)}_{k}\nonumber\\
&=\sqrt{\frac{\pr}{M\sum_{i=1}^{K}\beta_{i}}}\sum_{i=1}^{K}\g^{H}_{k}\g_{i}x_{j(i,t)}+n^{(t)}_{k},\label{eq:y_kt}
\end{align} 
respectively.

The $k$-th user knows its own transmitted signal $x_k$ (or $x_{j(k-t,t)}$), so it can remove the self-interference prior to decoding. The received signal after self-interference cancelation is
\begin{align}
\tilde{y}^{(t)}_{k}
&=\sqrt{\frac{\pr}{M\sum_{i=1}^{K}\beta_{i}}}\|\g_{k}\|^2x_{j(k,t)}\nonumber\\
&+\sqrt{\frac{\pr}{M\sum_{i=1}^{K}\beta_{i}}}\!\!\sum\limits_{\underset{j(i,t) \neq j(k,t), j(k-t,t)}{i=1}}^{K}\!\!\!\!\!\!\!\!\g^{H}_{k}\g_{i}x_{j(i,t)}+n^{(t)}_{k}\label{eq:y_tilde_t}.
\end{align}
The corresponding downlink spectral efficiency for the $t$-th time-slot is 
\begin{align}\label{eq:R_dl_tilde_kt0}
&{\tt{R}}^{\text{dl},(t)}_{k}\nonumber\\
&=\mathbb{E}\left\{\log_{2}\left(1+\frac{\frac{\pr}{M\sum_{i=1}^{K}\beta_{i}}\|\g_{k}\|^4}{\frac{\pr}{M\sum_{i=1}^{K}\beta_{i}}\!\!\!\!\!\!\!\!\!\!\!\!\!\!\!\sum\limits_{\underset{j(i,t) \neq j(k,t), j(k-t,t)}{i=1}}^{K}\!\!\!\!\!\!\!\!\!\!\left|\g^{H}_{k}\g_{i}\right|^2+1}\right)\right\}.
\end{align}

\begin{proposition}\label{Pro:proposition_1}
The spectral efficiency ${\tt{R}}^{\text{dl},(t)}_{k}$ given by \eqref{eq:R_dl_tilde_kt0} can be lower bounded by		
	\begin{align}\label{eq:R_dl_tilde_g1o}
	&{{\tt{R}}}^{\text{dl},(t)}_{k}\geq\tilde{{\tt{R}}}^{\text{dl},(t)}_{k}\nonumber\\
	&=\log_{2}\!\left(1+\!\frac{\pr(M-1)(M-2)\beta_{k}^2}{\pr(M-2)\beta_{k}\!\!\!\!\!\!\!\!\!\!\!\sum\limits_{\underset{j(i,t)\neq j(k,t), j(k-t,t)}{i=1}}^{K}\!\!\!\!\!\!\!\!\!\!\!\!\beta_{i}+M\sum\limits_{i=1}^{K}\beta_{i}}\right).
	\end{align}
	\begin{IEEEproof}
		See Appendix~\ref{AP:proposition_2}.
	\end{IEEEproof}
\end{proposition}

\vspace{0.2cm}
\section{Multi-Way Transmission with Successive Cancelation Decoding}
In this section, we propose a novel transmission scheme which requires only $\lceil\frac{K-1}{2}\rceil +1$ time-slots for the information exchange among the $K$ users.

\subsection{Multiple-Access Phase}
The multiple-access phase is the same as the one of conventional transmission scheme. See Section~\ref{sec:conv_MA}.

\subsection{Broadcast Phase}
Here, we need only $\lceil\frac{K-1}{2}\rceil$ time-slots to transmit all $K$ symbols to all users. The main idea is that: at a given time-slot, the $k$-th user subtracts all symbols decoded in previous time-slots prior to decoding the desired symbol. Furthermore, after $\lceil\frac{K-1}{2}\rceil$ time-slots, user $k$ receives $\lceil\frac{K-1}{2}\rceil$ signals, and each signal is a linear combination of $K-\lceil\frac{K-1}{2}\rceil -1$ symbols. So it can detect all   $K-\lceil\frac{K-1}{2}\rceil -1$ symbols without any inter-user interference through the zero-forcing technique. A detailed presentation of the proposed scheme is now provided. 

1) First time-slot: The relay intends to send $x_{j(k,1)}$ to the $k$-th user, for $k=1, \ldots, K$. The signal vector transmitted from the relay is
\begin{align}\label{eq:s_1}
\B{s}^{(1)}=\sqrt{\frac{\pr}{M\sum_{i=1}^{K}\beta_{i}}}\sum_{i=1}^{K}\g_{i}x_{j(i,1)}.
\end{align}
Thus, the received signal at the $k$-th user is
\begin{align}\label{eq:y_k1}
y^{(1)}_{k}
&=\g^H_{k}\B{s}^{(1)}+n^{(1)}_{k}\nonumber\\
&=\sqrt{\frac{\pr}{M\sum_{i=1}^{K}\beta_{i}}}\sum_{i=1}^{K}\g^H_{k}\g_{i}x_{j(i,1)}+n^{(1)}_{k},
\end{align}	
where $n^{(1)}_{k}\sim\CN(0,1)$ is the additive noise at the $k$-th user in the first time-slot. Since user $k$ knows its transmitted signal $x_k$ (or $x_{j(k-1,1)}$), it can subtract the self-interference before detecting signal $x_{j(k,1)}$. Therefore, the received signal at user $k$ after self-interference cancelation is 
\begin{align}\label{eq:y_tilde_1}
\tilde{y}^{(1)}_{k}
&=\sqrt{\frac{\pr}{M\sum_{i=1}^{K}\beta_{i}}}\|\g_{k}\|^2x_{j(k,1)}\nonumber\\
&+\sqrt{\frac{\pr}{M\sum_{i=1}^{K}\beta_{i}}}\sum\limits_{\underset{j(i,1)\notin\V_{k,1}}{i=1}}^{K}\!\!\!\!\g^{H}_{k}\g_{i}x_{j(i,1)}+n^{(1)}_{k},
\end{align}
where 
\begin{align}\label{eq:V_kt}
\V_{k,t}\triangleq\{j(k-t,t),j(k-t+1,t),\dots,j(k,t)\}.
\end{align}

Then, the corresponding spectral efficiency is given by
\begin{align}
&{\tt{R}}^{\text{dl},(1)}_{k}\nonumber\\
&=\mathbb{E}\left\{\log_{2}\left(1+\frac{\frac{\pr}{M\sum_{i=1}^{K}\beta_{i}}\|\g_{k}\|^4}{\frac{\pr}{M\sum_{i=1}^{K}\beta_{i}}\!\!\!\!\sum\limits_{\underset{j(i,1)\notin\V_{k,1}}{i=1}}^{K}\!\!\!\!\left|\g^{H}_{k}\g_{i}\right|^2+1}\right)\right\}.
\end{align}

2) Second time-slot: The relay intends to send $x_{j(k,2)}$ to the $k$-th user, for $k=1, \ldots, K$. The signal vector transmitted from the relay is
\begin{align}\label{eq:s_2}
\B{s}^{(2)}=\sqrt{\frac{\pr}{M\sum_{i=1}^{K}\beta_{i}}}\sum_{i=1}^{K}\g_{i}x_{j(i,2)},
\end{align}
and hence, the  signal received at the $k$-th user is
\begin{align}
y^{(2)}_{k}
&=\g^H_{k}\B{s}^{(2)}+n^{(2)}_{k}\nonumber\\
&=\sqrt{\frac{\pr}{M\sum_{i=1}^{K}\beta_{i}}}\sum_{i=1}^{K}\g^{H}_{k}\g_{i}x_{j(i,2)}+n^{(2)}_{k}.
\end{align}

The $k$-th user knows its own transmitted symbol $x_k$ as well as the symbol detected in the first time-slot $x_{j(k,1)}$, so it can subtract these symbols before detecting the desired signal $x_{j(k,2)}$. The received signal at the $k$-th user after subtracting the above symbols is 
\begin{align}\label{eq:y_tilde_2}
\tilde{y}^{(2)}_{k}
&=\sqrt{\frac{\pr}{M\sum_{i=1}^{K}\beta_{i}}}\|\g_{k}\|^2x_{j(k,2)}\nonumber\\
&+\sqrt{\frac{\pr}{M\sum_{i=1}^{K}\beta_{i}}}\!\!\sum\limits_{\underset{j(i,2)\notin\V_{k,2}}{i=1}}^{K}\!\!\!\!\g^{H}_{k}\g_{i}x_{j(i,2)}+n^{(2)}_{k}.
\end{align}

Then, the spectral efficiency of user $k$ at the second time-slot is
\begin{align}
&{\tt{R}}^{\text{dl},(2)}_{k}\nonumber\\
&=\mathbb{E}\left\{\log_{2}\left(\!1\!+\frac{\frac{\pr}{M\sum_{i=1}^{K}\beta_{i}}\|\g_{k}\|^4}{\frac{\pr}{M\sum_{i=1}^{K}\beta_{i}}\!\!\!\!\!\!\!\sum\limits_{\underset{j(i,2)\notin\V_{k,2}}{i=1}}^{K}\!\!\!\!\!\!\!\left|\g^{H}_{k}\g_{i}\right|^2\!+\!1}\right)\right\}.
\end{align}

3) $t$-th time-slot: At the $t$-time-slot, the relay intends to send $x_{j(k,t)}$ to the $k$-th user, for $k=1, \ldots, K$. The signal vector transmitted from the relay is
\begin{align}
\B{s}^{(t)}
&=\sqrt{\frac{\pr}{M\sum_{i=1}^{K}\beta_{i}}}\sum_{i=1}^{K}\g_{i}x_{j(i,t)}.\label{eq:s_t}
\end{align}
Then, the $k$-th user sees
\begin{align}
y^{(t)}_{k}
&=\g^H_{k}\B{s}^{(t)}+n^{(t)}_{k}\nonumber\\
&=\sqrt{\frac{\pr}{M\sum_{i=1}^{K}\beta_{i}}}\sum_{i=1}^{K}\g^{H}_{k}\g_{i}x_{j(i,t)}+n^{(t)}_{k}.\label{eq:y_kt}
\end{align} 

The $k$-th users know its own transmitted symbols $x_k$. Furthermore, it also knows its detected symbols in previous time-slots. So it knows $\{x_{j(k-1,1)},x_{j(k,1)}, x_{j(k,2)},\dots, x_{j(k,t-1)}\}$, and, hence, it can remove these symbols to obtain
\begin{align}
\tilde{y}^{(t)}_{k}
&=\sqrt{\frac{\pr}{M\sum_{i=1}^{K}\beta_{i}}}\|\g_{k}\|^2x_{j(k,t)}\nonumber\\
&+\sqrt{\frac{\pr}{M\sum_{i=1}^{K}\beta_{i}}}\!\!\sum\limits_{\underset{j(i,t)\notin\V_{k,t}}{i=1}}^{K}\!\!\!\!\g^{H}_{k}\g_{i}x_{j(i,t)}+n^{(t)}_{k}\label{eq:y_tilde_t}.
\end{align}

Then, the spectral efficiency of the $k$-th user at the $t$-th time-slot is 
\begin{align}\label{eq:R_dl_tilde_kt}
&{\tt{R}}^{\text{dl},(t)}_{k}\nonumber\\
&=\mathbb{E}\left\{\log_{2}\left(1+\frac{\frac{\pr}{M\sum_{i=1}^{K}\beta_{i}}\|\g_{k}\|^4}{\frac{\pr}{M\sum_{i=1}^{K}\beta_{i}}\!\!\!\!\!\!\!\sum\limits_{\underset{j(i,t)\notin\V_{k,t}}{i=1}}^{K}\!\!\!\!\!\!\!\left|\g^{H}_{k}\g_{i}\right|^2\!+\!1}\right)\right\}\!.
\end{align}
\begin{proposition}\label{Pro:proposition_2}
	The spectral efficiency ${\tt{R}}^{\text{dl},(t)}_{k}$ given by \eqref{eq:R_dl_tilde_kt} can be lower bounded by
	\begin{align}\label{eq:R_dl_tilde_g1}
	&{\tt{R}}^{\text{dl},(t)}_{k}\geq\tilde{{\tt{R}}}^{\text{dl},(t)}_{k}\nonumber\\
	&=\log_{2}\!\left(1+\!\frac{\pr(M-1)(M-2)\beta_{k}^2}{\pr(M-2)\beta_{k}\!\!\!\!\!\sum\limits_{\underset{j(i,t)\notin\V_{k,t}}{i=1}}^{K}\!\!\!\!\!\!\beta_{i}+M\sum\limits_{i=1}^{K}\beta_{i}}\right).
	\end{align}
	
	\begin{IEEEproof}
		Following a similar methodology as the proof of Proposition~\ref{Pro:proposition_1}.
	\end{IEEEproof}
\end{proposition}

4) After $t'=\left \lceil \frac{K-1}{2} \right \rceil$ time-slots, the $k$-th user has received  $t'$ signals (the $t$-th received signal is given by \eqref{eq:y_kt}). Furthermore, it has decoded $t'$ symbols. So it can subtract all $t'$ detected symbols from each received signal to obtain the following results:
\begin{align}\label{eq:linear_equations}
\left\{\begin{matrix}
\bar{y}^{(t')}_{k,1}\!\!\!\!\!\!\!
&=&\!\!\!\!\!\!\!\sqrt{\frac{\pr}{M\sum_{i=1}^{K}\beta_{i}}}\!\!\!\!\!\!\!\sum\limits_{\underset{j(i,t')\notin\Vt}{i=1}}^{K}\!\!\!\!\!\!\g^H_{k}\g_{j(k,1)}x_{j(i,t')} + n^{(t')}_{k,1},
&   \\ 
\bar{y}^{(t')}_{k,2}\!\!\!\!\!\!\!
&=&\!\!\!\!\!\!\!\sqrt{\frac{\pr}{M\sum_{i=1}^{K}\beta_{i}}}\!\!\!\!\!\!\!\sum\limits_{\underset{j(i,t')\notin\Vt}{i=1}}^{K}\!\!\!\!\!\!\g^H_{k}\g_{j(k,2)}x_{j(i,t')} + n^{(t')}_{k,2}, 
&   \\ 
\vdots 
&   \\
\bar{y}^{(t')}_{k,t'}\!\!\!\!\!\!\!\!
&=&\!\!\!\!\sqrt{\frac{\pr}{M\sum_{i=1}^{K}\beta_{i}}}\!\!\!\!\!\!\!\sum\limits_{\underset{j(i,t')\notin\Vt}{i=1}}^{K}\!\!\!\!\!\!\!\g^H_{k}\g_{j(k,t')}x_{j(i,t')} + n^{(t')}_{k,t'}.   
\end{matrix}\right.
\end{align}
We can see that we have $t'$ equations, each equation has $(K-t'-1)$ unknown variables $\{x_{j(i,t')}\}$. Since $t'=\left \lceil \frac{K-1}{2} \right \rceil$, the number of equations is greater than or equal to the number of unknown variables. Therefore, the $k$-th user can detect all remaining $(K-t'-1)$ symbols $\{x_{j(i,t')}\}$ via the ZF scheme as follows. Denote by
\begin{align}\label{eq:y,n_bar}
\bar{\B{y}}^{(t')}_{k}
&\triangleq
\begin{bmatrix}
\bar{y}^{(t')}_{k,1}
\\ 
\bar{y}^{(t')}_{k,2}
\\
\vdots 
\\
\bar{y}^{(t')}_{k,t'}
\end{bmatrix}, \ \
\bar{\B{n}}^{(t')}_{k}
\triangleq
\begin{bmatrix}
n^{(t')}_{k,1}
\\ 
n^{(t')}_{k,2}
\\
\vdots 
\\
n^{(t')}_{k,t'}
\end{bmatrix},
\end{align}
\begin{align}\label{eq:A}
\B{A}_{k}
\!\triangleq\!\!
\begin{bmatrix}
\g^H_{k}\g_{j(k,1)}   	&\!\!\!\!\!\!\!\!\!\!\!\!\!\!\g^H_{k}\g_{j(k,2)}\!\!\!\!\!\!\!\!		&\dots 		&\!\g^H_{k}\g_{j(k,K-t'-1)}
\\ 
\g^H_{k}\g_{j(k,2)} 	&\!\!\!\!\!\!\!\!\!\!\g^H_{k}\g_{j(k,3)}\!\!\!\! 		&\dots 	&\!\!\!\!\!\!\!\g^H_{k}\g_{j(k,K-t')}
\\ 
\vdots 			 		&\vdots						& 			&\vdots
\\ 
\g^H_{k}\g_{j(k,t')}  	&\g^H_{k}\g_{j(k,t'+1)}  	&\dots 		&\!\!\!\!\!\!\!\g^H_{k}\g_{j(k,K-2)}
\end{bmatrix},
\end{align}
and
\begin{align}\label{eq:x_bar}
\bar{\B{x}}
\triangleq
\begin{bmatrix}
x_{j(k,t'+1)} & x_{j(k,t'+2)} &\dots & x_{j(k,K-1)}
\end{bmatrix}^T.
\end{align}
Then, (\ref{eq:linear_equations}) can be rewritten in matrix-vector form as
\begin{align}\label{eq:y_bar}
\bar{\B{y}}^{(t')}_{k}
=\sqrt{\frac{\pr}{M\sum_{i=1}^{K}\beta_{i}}}\B{A}_{k}\bar{\B{x}}+\bar{\B{n}}^{(t')}_{k}.
\end{align}

The $k$-th user applies the ZF scheme to decode the remaining symbols as follows:
\begin{align}\label{eq:zfnew1}
\tilde{\B{r}}^{(t')}_{k}
=\B{Z}^T\bar{\B{y}}^{(t')}_{k}
=\sqrt{\frac{\pr}{M\sum_{i=1}^{K}\beta_{i}}}\B{Z}^T\B{A}_{k}\bar{\B{x}}+\B{Z}^T\bar{\B{n}}^{(t')}_{k},
\end{align}
where
\begin{align}\label{eq:Z^T}
\B{Z}^T\triangleq\left(\B{A}_{k}^H\B{A}_{k}\right)^{-1}\B{A}_{k}^H.
\end{align}

The $n$-th element of $\tilde{\B{r}}^{(t')}_{k}$ will be used to detect $x_{j(k,t'+n)}$. From \eqref{eq:zfnew1} and the fact that $\B{Z}^T\B{A}_{k}=\B{I}_{K-(t'+1)}$, the $n$-th element of $\tilde{\B{r}}^{(t')}_{k}$ is given by
\begin{align}\label{eq:r_tilde_kk'}
\tilde{{\text{r}}}_{k,n}^{(t')}
&=\sqrt{\frac{\pr}{M\sum_{i=1}^{K}\beta_{i}}}x_{j(k,t'+n)}+\B{z}^T_{n}\bar{\B{n}}^{(t')}_{k}.
\end{align}

Thus, the corresponding spectral efficiency of the system in \eqref{eq:r_tilde_kk'} is
\begin{align}\label{eq:R_tilde_dl_kk'}
&{\tt{R}}^{\text{dl},(t'+n)}_{k}
=\mathbb{E}\left\{\log_{2}\left(1+\frac{\frac{\pr}{M\sum_{i=1}^{K}\beta_{i}}}{\|\B{z}_{n}\|^2}\right)\right\}\nonumber\\
&=\mathbb{E}\left\{\log_{2}\left(1+\frac{\pr}{M\sum_{i=1}^{K}\beta_{i}\left[\left(\B{A}_{k}^H\B{A}_{k}\right)^{-1}\right]_{nn}}\right)\right\}.
\end{align}

Since \eqref{eq:R_tilde_dl_kk'} has a complicated form that involves a matrix inverse, we cannot obtain an exact closed-form. However, thanks
to the trace lemma and law of large numbers (as $M$ goes to infinity) \cite{wagner2012large}, we can obtain the
following approximating result.

\begin{proposition}\label{Pro:proposition_3}
As  $M\to\infty$, the spectral efficiency ${\tt{R}}^{\text{dl},(t'+n)}_{k}$ given by \eqref{eq:R_tilde_dl_kk'} converges to
	\begin{align}\label{eq:R_tilde_dl_kk1}
	{\tt{R}}^{\text{dl},(t'+n)}_{k}\to\log_{2}\left(1+\frac{\pr\beta_{k}\sum_{i=1}^{t'}\beta_{j(k,n+i-1)}}{\sum_{i=1}^{K}\beta_{i}}\right).
	\end{align}

	\begin{IEEEproof}
		See Appendix~\ref{AP:proposition_3}.
	\end{IEEEproof}
\end{proposition}

\vspace{0.3cm}
\section{Numerical Results}

In this section, we provide numerical results to evaluate the performance of our proposed scheme. We consider the sum spectral efficiency, defined as
\begin{align}
{\tt{SE}}_{\text{sum}}=\frac{1}{t'+1}\sum_{k=1}^{K}\sum_{t=1}^{K-1}\text{min}\left({\tt{R}}^{\text{ul}}_{k},{\tt{R}}^{\text{dl},(t)}_{k}\right),
\end{align}
where $t'=\left \lceil \frac{K-1}{2} \right \rceil$ is the $t'$-th time-slot of the transmission protocol in broadcast phase.

\begin{figure}[t]
	\begin{center}
		\includegraphics[scale=0.24]{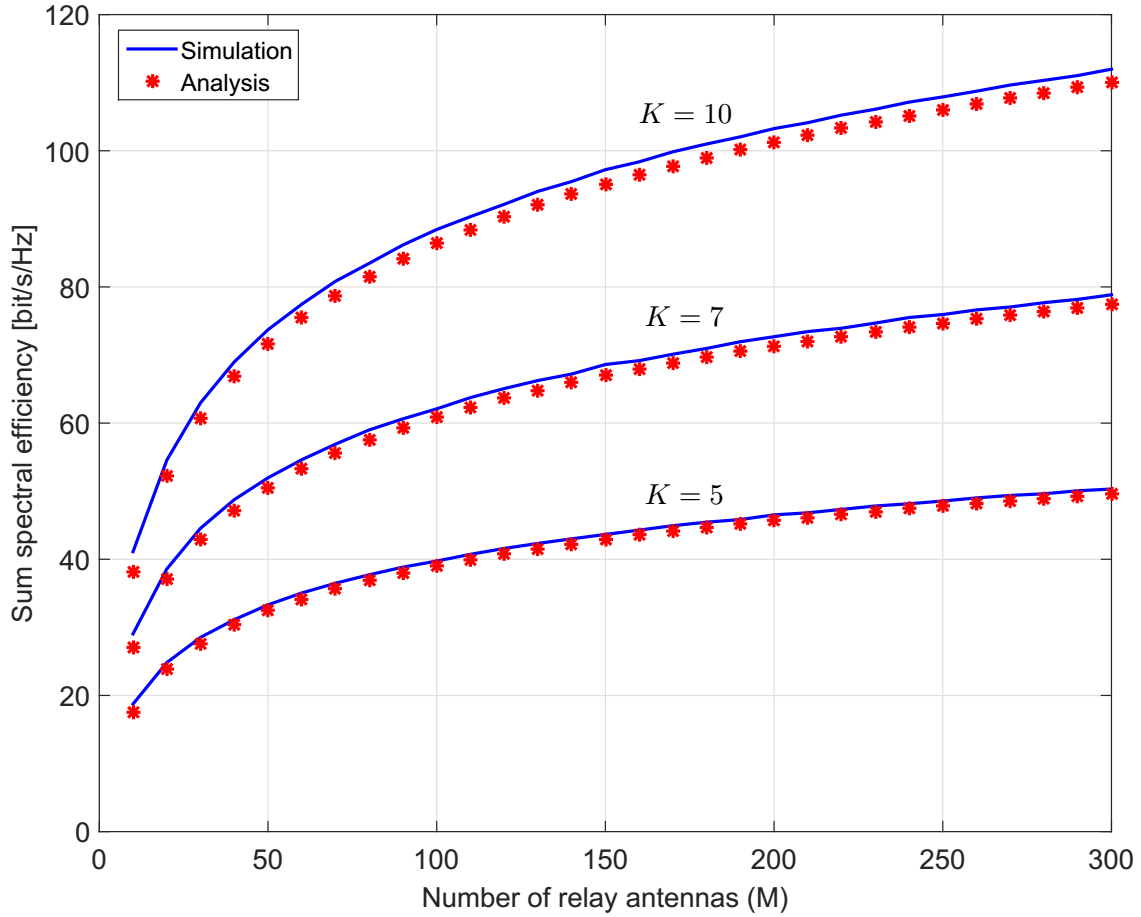}		
	\end{center}
	\caption{The sum spectral efficiency of the system model with different $K$ versus the number of relay antennas. We set
		$\pu=0$ dB, $\pr=10$ dB, $\beta_{k}=1$.
	}\label{fig:SE_group12}
\end{figure}

First, we examine the tightness of our analytical results. Figure~\ref{fig:SE_group12} shows the sum spectral efficiency of our proposed scheme versus the number of relay antennas with different $K$ for the simple case $\beta_{k}=1$, $\forall k$. The ``analysis'' curves represent our analytical results obtained by using the lower bounds \eqref{eq:R_tilde_ul}, \eqref{eq:R_dl_tilde_g1}, and the asymptotic result \eqref{eq:R_tilde_dl_kk1}. The ``simulation'' curves are generated from the outputs of a Monte-Carlo simulator using \eqref{eq:R_ul}, \eqref{eq:R_dl_tilde_kt}, and \eqref{eq:R_tilde_dl_kk'}. We can see that the  proposed  approximation is  very  tight, even with small number of antennas. Furthermore, as expected, the sum spectral efficiency increases significantly when the number of relay antennas increases. 

\begin{figure}[t]
	\begin{center}
		\includegraphics[scale=0.24]{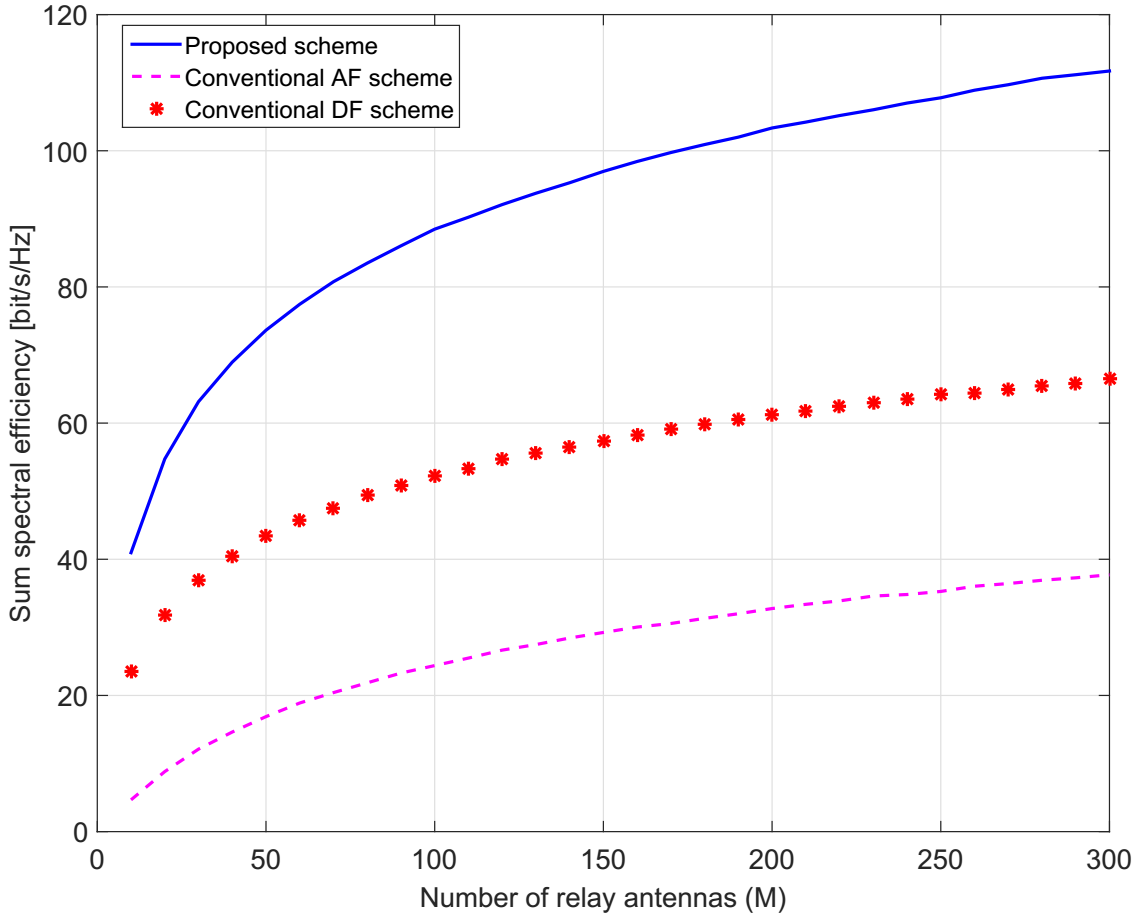}		
	\end{center}
	\caption{The comparison of the sum spectral efficiency with different schemes versus the number of relay antennas. We choose $\pu=0$ dB, $\pr=10$ dB, $K=10$, $\beta_{k}=1$.
	}\label{fig:SE_diff_PS_CS_Sigte}
\end{figure}

We next compare the performance of our proposed scheme with the one of the conventional DF scheme (Section~\ref{sec:conven_broads}) and the conventional AF scheme in \cite{ho2017multiway} (see Fig.~\ref{fig:SE_diff_PS_CS_Sigte}). We can see that our proposed scheme significantly outperforms other schemes. The sum spectral efficiency of our proposed scheme improves by factors of nearly $2$ and $3$ compared with the conventional DF scheme and the conventional AF scheme, respectively. This is due to the fact that with the conventional DF scheme, we need in total $K$ time-slots to exchange the information among the $K$ users, while with our proposed scheme, we need only  $\left \lceil \frac{K-1}{2} \right \rceil+1$.

\begin{figure}[t]
	\begin{center}
		\includegraphics[scale=0.24]{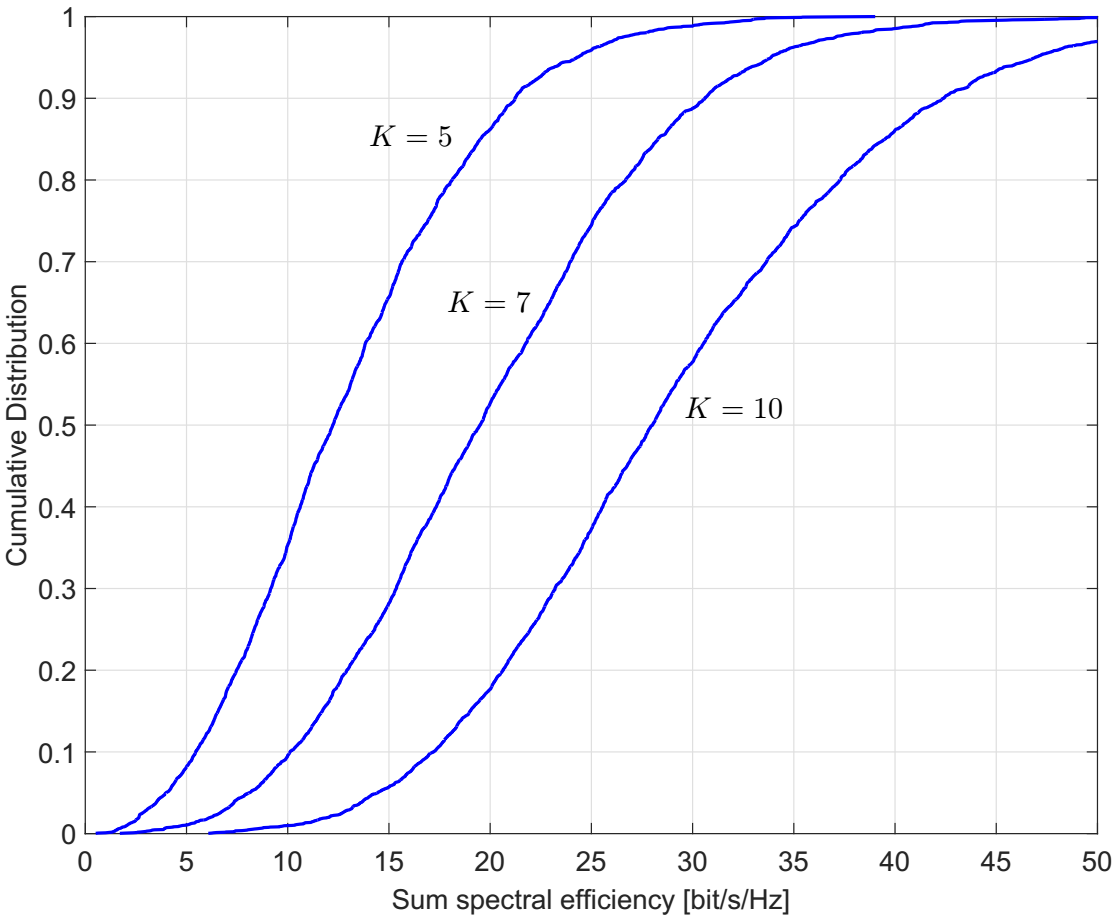}		
	\end{center}
	\caption{Cumulative distribution of the sum spectral efficiency for different $K$. We choose
		$\pu=0$ dB, $\pr=10$ dB, $M=100$.
	}\label{fig:Cumulative}
\end{figure}

Finally, we consider a more practical scenario where the  large-scale fading $\beta_k$ changes depending on the locations of users and the shadow fading. To generate the large-scale fading, we use the same model as in \cite{ho2017performance}. Figure~\ref{fig:Cumulative} illustrates the cumulative distribution of the sum spectral efficiency of our proposed scheme for $K=5, 7$, and $10$. As expected, the sum spectral efficiency increases when $K$ increases. The 95\%-likely sum spectral efficiency with $K=10$ is about $14.5$ bit/s/Hz which is nearly 4 times and 2 times higher than that with $K=5$ and $K=7$, respectively.

\vspace{0.3cm}
\section{Conclusion}
We proposed a novel and useful transmission scheme for multi-way massive MIMO  relay systems with decode-and-forward protocol at the relay. While the conventional scheme needs $K$ time-slots to exchange all data among $K$ users, our proposed scheme, which is based on successive cancelation decoding, needs only  
 $\left \lceil \frac{K-1}{2} \right \rceil+1$ time-slots. Thus, the sum spectral efficiency of our proposed scheme is nearly double the sum spectral efficiency of the conventional scheme.

\vspace{0.3cm}
\section{Appendices}

\subsection{Proof of Proposition~\ref{Pro:proposition_1}}\label{AP:proposition_2}

By using Jensen's inequality, we obtain 
	\begin{align}\label{eq:R_tilde_dl_g1}
	&{\tt{R}}^{\text{dl},(t)}_{k}\geq \tilde{{\tt{R}}}^{\text{dl},(t)}_{k}=\log_{2}\!\left(\!\!1+\!\!\left(\mathbb{E}\left\{\mathcal{X}_{k}^{(t)}\right\}\right)^{-1}\right),
	\end{align}
where 
	\begin{align}\label{eq:R_tilde_dl_g1x}
	\mathcal{X}_{k}^{(t)} \triangleq 
	\frac{\frac{\pr}{M\sum_{i=1}^{K}\beta_{i}}\!\!\!\!\!\!\!\!\!\!\sum\limits_{\underset{j(i,t) \neq j(k,t), j(k-t,t)}{i=1}}^{K}\!\!\!\!\!\!\!\!\!\!\!\left|\g^{H}_{k}\g_{i}\right|^2+1}{\frac{\pr}{M\sum_{i=1}^{K}\beta_{i}}\|\g_{k}\|^4}.
	\end{align}
By dividing the numerator and the denominator of the right-hand side of \eqref{eq:R_tilde_dl_g1x} by $\|\g_{k}\|^2$, we get
	\begin{align}\label{eq:R_tilde_dl_g1x1}
	\mathbb{E}\!\left\{\mathcal{X}_{k}^{(t)}\right\}\!
	=\!\mathbb{E}\left\{\frac{\!\sum\limits_{\underset{j(i,t)\neq j(k,t), j(k-t,t)}{i=1}}^{K}\!\!\!\!\!\!\!\!\!\!\!\left|\tilde{g}_{i}\right|^2}{\|\g_{k}\|^2}\!+\!\frac{1}{\frac{\pr}{M\sum_{i=1}^{K}\beta_{i}}\|\g_{k}\|^4}\right\}\!,
	\end{align}
where $\tilde{g}_{i}\triangleq \frac{\left|\g^{H}_{k}\g_{i}\right|^2}{\|\g_{k}\|^2}$. Conditioned on $\g_{k}$, $\tilde{g}_{i}$ is Gaussian distributed with zero mean and variance $\beta_i$. Since the variance of $\tilde{g}_{i}$ does not depend on $\g_{k}$, $\tilde{g}_{i}$ is a $\CN(0,\beta_i)$ random variable and  is independent of  $\g_{k}$. Therefore, 
	\begin{align}\label{eq:R_tilde_dl_g1x1}
	\mathbb{E}\left\{\mathcal{X}_{k}^{(t)}\right\}
	&=\hspace{-0.5cm}
	\sum\limits_{\underset{j(i,t)\neq j(k,t), j(k-t,t)}{i=1}}^{K}\hspace{-0.8cm}\mathbb{E}\left\{\left|\tilde{g}_{i}\right|^2\right\}\mathbb{E}\left\{\frac{1}{\|\g_{k}\|^2}\right\}\nonumber\\
	&+\mathbb{E}\left\{\frac{1}{\frac{\pr}{M\sum_{i=1}^{K}\beta_{i}}\|\g_{k}\|^4}\right\}.
	\end{align}

By using \cite[Lemma 2.10]{tulino2004random}, we obtain $\mathbb{E}\left\{\frac{1}{\|\g_{k}\|^2}\right\} = \frac{1}{M-1}$ and  $\mathbb{E}\left\{\frac{1}{\|\g_{k}\|^4}\right\} = \frac{M}{(M-1)^3 - (M-1)}$, and hence, we arrive at \eqref{eq:R_dl_tilde_g1o}.

\subsection{Proof of Proposition~\ref{Pro:proposition_3}}\label{AP:proposition_3}

From \eqref{eq:A}, the $(m,n)$-th element of $\B{A}_{k}^H\B{A}_{k}$ is given by
\begin{align}\label{eq:A^HA}
\left[\B{A}_{k}^H\B{A}_{k}\right]_{mn} = \sum\limits_{i=1}^{t'}\g^H_{k}\g_{j(k,i-1+m)}\g_{j(k,i-1+n)}^H\g_{k}.
\end{align}
 Using the trace lemma \cite[Lemmas 4, 5]{wagner2012large}, we have
\begin{align}\label{eq:trc1x}
\frac{1}{M}\g^H_{k}\g_{j(k,i)}\g_{j(k,i)}^H\g_{k}  - \frac{\beta_k}{M}\text{Tr}\left(\g_{j(k,i)}\g_{j(k,i)}^H\right)\xrightarrow[M\to\infty]{a.s.}0,
\end{align}
where ``a.s.'' stands for almost sure convergence.

Since $\text{Tr}\left(\g_{j(k,i)}\g_{j(k,i)}^H\right) = \|\g_{j(k,i)}\|^2$, and from the law of large numbers, we get
\begin{align}\label{eq:trc2x}
 \frac{1}{M}\text{Tr}\left(\g_{j(k,i)}\g_{j(k,i)}^H\right)\xrightarrow[M\to\infty]{a.s.} \beta_{j(k,i)}.
\end{align}
The substitution of \eqref{eq:trc2x} into \eqref{eq:trc1x} yields
\begin{align}\label{eq:trc1}
\frac{1}{M}\g^H_{k}\g_{j(k,i)}\g_{j(k,i)}^H\g_{k}\xrightarrow[M\to\infty]{a.s.} \beta_k \beta_{j(k,i)}.
\end{align}

Similarly, we obtain
\begin{align}\label{eq:trc2}
\frac{1}{M}\g_{j(k,i)}^H\g_{k}\g^H_{k}\g_{j(k,i+1)}  \xrightarrow[M\to\infty]{a.s.} 0.
\end{align}

From \eqref{eq:A^HA}, \eqref{eq:trc1}, and \eqref{eq:trc2}, we have
\begin{align}\label{eq:E_AA}
M\left[\left(\B{A}_{k}^H\B{A}_{k}\right)^{-1}\right]_{nn}
\xrightarrow[]{a.s.} \frac{1}{\beta_{k}\sum_{i=1}^{t'}\beta_{j(k,n+i-1)}}.
\end{align}

Substituting (\ref{eq:E_AA}) into (\ref{eq:R_tilde_dl_kk'}), we obtain (\ref{eq:R_tilde_dl_kk1}).

\vspace{0.2cm}
\section*{Acknowledgment}
This work was supported by project no. 3811/QD-UBND, Binh Duong government, Vietnam. The work of H. Q. Ngo was supported by the Swedish Research Council (VR) and ELLIIT. The work of M. Matthaiou was supported in part by the EPSRC under grant EP/P000673/1. The work of T. Q. Duong was supported by  the U.K. Royal Academy of Engineering Research Fellowship under Grant RF1415$\backslash$14$\backslash$22, and by the EPSRC under Grant EP/P019374/1.

\bibliographystyle{IEEEtran}
\bibliography{bibfile_gen}

\begin{thebibliography}{10}
\providecommand{\url}[1]{#1}
\csname url@samestyle\endcsname
\providecommand{\newblock}{\relax}
\providecommand{\bibinfo}[2]{#2}
\providecommand{\BIBentrySTDinterwordspacing}{\spaceskip=0pt\relax}
\providecommand{\BIBentryALTinterwordstretchfactor}{4}
\providecommand{\BIBentryALTinterwordspacing}{\spaceskip=\fontdimen2\font plus
\BIBentryALTinterwordstretchfactor\fontdimen3\font minus
  \fontdimen4\font\relax}
\providecommand{\BIBforeignlanguage}[2]{{%
\expandafter\ifx\csname l@#1\endcsname\relax
\typeout{** WARNING: IEEEtran.bst: No hyphenation pattern has been}%
\typeout{** loaded for the language `#1'. Using the pattern for}%
\typeout{** the default language instead.}%
\else
\language=\csname l@#1\endcsname
\fi
#2}}
\providecommand{\BIBdecl}{\relax}
\BIBdecl

\bibitem{ho2017performance}
C.~D. Ho, H.~Q. Ngo, M.~Matthaiou, and T.~Q. Duong, ``On the performance of
  zero-forcing processing in multi-way massive {MIMO} relay networks,''
  \emph{to appear IEEE Commun. Letters}, 2017.

\bibitem{larsson2014massive}
E.~G. Larsson, O.~Edfors, F.~Tufvesson, and T.~L. Marzetta, ``Massive {MIMO}
  for next generation wireless systems,'' \emph{IEEE Commun. Mag.}, vol.~52,
  no.~2, pp. 186--195, Feb. 2014.

\bibitem{ngo2013energy}
H.~Q. Ngo, E.~G. Larsson, and T.~L. Marzetta, ``Energy and spectral efficiency
  of very large multiuser {MIMO} systems,'' \emph{IEEE Trans. Commun.},
  vol.~61, no.~4, pp. 1436--1449, Apr. 2013.

\bibitem{marzetta2016fundamentals}
T.~L. Marzetta, E.~G. Larsson, H.~Yang, and H.~Q. Ngo, \emph{Fundamentals of
  Massive MIMO}.\hskip 1em plus 0.5em minus 0.4em\relax Cambridge University
  Press, 2016.

\bibitem{gunduz2013multiway}
D.~G{\"u}nd{\"u}z, A.~Yener, A.~Goldsmith, and H.~V. Poor, ``The multiway relay
  channel,'' \emph{IEEE Trans. Inf. Theory}, vol.~59, no.~1, pp. 51--63, Jan.
  2013.

\bibitem{tian2014degrees}
Y.~Tian and A.~Yener, ``Degrees of freedom for the {MIMO} multi-way relay
  channel,'' \emph{IEEE Trans. Inf. Theory}, vol.~60, no.~5, pp. 2495--2511,
  May 2014.

\bibitem{amah2009non}
A.~Amah and A.~Klein, ``Non-regenerative multi-way relaying with linear
  beamforming.'' in \emph{Proc. IEEE PIMRC}, Sep. 2009, pp. 1843--1847.

\bibitem{badugewireless}
G.~Amarasuriya, E.~G. Larsson, and H.~V. Poor, ``Wireless information and power
  transfer in multi-way massive {MIMO} relay networks,'' \emph{IEEE Trans.
  Wireless Commun.}, vol.~15, no.~6, pp. 3837--3855, June 2015.

\bibitem{amarasuriya2014multi}
G.~Amarasuriya and H.~V. Poor, ``Multi-way amplify-and-forward relay networks
  with massive {MIMO},'' in \emph{Proc. IEEE PIMRC}, Sep. 2014, pp. 595--600.

\bibitem{wagner2012large}
S.~Wagner, R.~Couillet, M.~Debbah, and D.~T. Slock, ``Large system analysis of
  linear precoding in {MISO} broadcast channels with limited feedback,''
  \emph{IEEE Trans. Inf. Theory}, vol.~58, no.~7, pp. 4509--4537, July 2012.

\bibitem{ho2017multiway}
C.~Ho, H.~Q. Ngo, M.~Matthaiou, and T.~Q. Duong, ``Multi-way massive {MIMO}
  relay networks with maximum-ratio processing,'' in \emph{Proc. IEEE
  SigTelCom}, Jan. 2017, pp. 124--128.

\bibitem{tulino2004random}
A.~M. Tulino and S.~Verd{\'u}, ``Random matrix theory and wireless
  communications,'' \emph{Foundations and Trends in Commun. and Inf. Theory},
  vol.~1, no.~1, pp. 1--182, Jun. 2004.

\end{thebibliography}

\end{document}